\documentclass[journal]{IEEEtran}

\usepackage[pdftex]{graphicx}
\DeclareGraphicsExtensions{.pdf,.jpeg,.png}

\usepackage[cmex10]{amsmath}
\usepackage{amsfonts}
\usepackage{cite}
\usepackage{url}
\usepackage{color}
\graphicspath{{figures/}}

\newcommand{\BE}{\begin{equation}}
\newcommand{\EE}{\end{equation}}
\newcommand{\BF}{\begin{figure}\centering}
\newcommand{\EF}{\end{figure}}
\newcommand{\BT}{\begin{table}\centering}
\newcommand{\ET}{\end{table}}
\newcommand{\RE}[1]{\Re\left\{ #1 \right\}}
\newcommand{\IM}[1]{\Im\left\{ #1 \right\}}

\newcommand{\Qc}{Q_\mathrm{cl}}
\newcommand{\Qrev}{Q_\mathrm{rev}}
\newcommand{\QX}{Q_X}
\newcommand{\QZ}{Q_Z}
\newcommand{\Wst}{W_\mathrm{sto}}
\newcommand{\Wrev}{W_\mathrm{rev}}
\newcommand{\PL}{P_\mathrm{lost}}
\newcommand{\Zin}{Z_\mathrm{in}}
 
\begin{document}
\title{On the Functional Relation between Quality Factor and Fractional Bandwidth}
\author{Miloslav~Capek,~\IEEEmembership{Member,~IEEE,}
        Lukas~Jelinek,
        and~Pavel~Hazdra,~\IEEEmembership{Member,~IEEE}
\thanks{Manuscript received March 20, 2014; revised March 20, 2014.
This work was supported by the Czech Science Foundation under project 13-09086S and by the COST IC1102 (VISTA) action.}
\thanks{The authors are with the Department of Electromagnetic Field, Faculty of Electrical Engineering, Czech Technical University in Prague, Technicka 2, 16627, Prague, Czech Republic
(e-mail: jelinel1@fel.cvut.cz).}% <-this % stops a space
}

\markboth{Journal of \LaTeX\ Class Files,~Vol.~6, No.~1, January~2007}%
{Authors \MakeLowercase{\textit{et al.}}: On the Relationship Between Quality Factor and Fractional Bandwidth}
% The only time the second header will appear is for the odd numbered pages after the title page when using the twoside option.
\maketitle

\begin{abstract}
The functional relation between the fractional bandwidth and the quality factor of a radiating system is investigated in this note. Several widely used definitions of the quality factor are compared on two examples of RLC circuits that serve as a simplified model of a single resonant antenna tuned to its resonance. It is demonstrated that for a first-order system, only the quality factor based on differentiation of input impedance has unique proportionality to the fractional bandwidth, whereas e.g. the classical definition of the quality factor, i.~e. the ratio of the stored energy to the lost energy per one cycle, is not uniquely proportional to the fractional bandwidth. In addition, it is shown that for higher-order systems the quality factor based on differentiation of the input impedance ceases to be uniquely related to the fractional bandwidth.
\end{abstract}


% \begin{center} \mathbfseries EDICS Category: 3-BBND \end{center}
% \fi
%
% For peerreview papers, this IEEEtran command inserts a page break and creates the second title. It will be ignored for other modes.
\IEEEpeerreviewmaketitle

%%%%%%%%%%%%%%%%%%%%%%%%%%%%%%%%%%%%%%%%%%%%%%%%%%%%%%%%%%%%%%%%%%%%%%%%%%%%%%%%%%%%%%%%%%%%%%%%%%%%%%%%%%%%%%%%%%%%%%%%%%%%%%%%%%%%%%%%%%%%%%%%%%%%%%%%%%%%%%%%%%%%%%%%%%%%%%%%
\section{Introduction}
\label{Intro}
\IEEEPARstart{T}{he} fractional bandwidth (FBW) is a parameter of primary importance in any oscillating system \cite{MorseFeshBach_MethodsOfTheoreticalPhysics}, since it is a relative frequency band in which the system can be effectively driven by an external source. In the case of an antenna, a fractional bandwidth is a frequency band in which the power incident upon the input port can be effectively radiated \cite{Balanis_Wiley_2005}.

Based on an analytical evaluation of the basic RLC circuits in the time-harmonic domain \cite{Hallen_ElectromagneticTheory}, the FBW is believed to be inversely proportional to the quality factor, which is commonly defined as $2 \pi$ times the ratio of the cycle mean stored energy and the lost energy, see e.g. IEEE Std. 145-1993, \cite{IEEEStd_antennas}. This relation is known to be very precise for high values of Q, and has been shown to be exact for Q tending to infinity, i.e. a lossless oscillating system cannot be driven by an external source, since its FBW is equal to zero. However, this inverse proportionality is known to fail at low values of Q and, in fact, it is not clear whether there exists any functional relation of FBW and Q which would be valid in all ranges of Q. It is however important to stress that if such a relation were to exist, it would be of crucial importance, since there exists a fundamental lower bound of Q of a lossless electromagnetic radiator \cite{Chu_PhysicalLimitationsOfOmniDirectAntennas, HansenCollin_ANewChuFormulaForQ}, which would then imply a fundamental upper bound of its FBW, an essential theoretical limitation for electrically small radiators.
%Nevertheless, it will be demonstrated that this Q definition has no unambiguous proportionality to the FBW. In an antenna theory, the exact relation between FBW and Q would however be of crucial importance, since there exist lower bounds of Q, which would imply the upper bounds of FBW, an essential theoretical limitation for electrically small antennas.

This note serves two purposes. First, a proof is given of the non-existence of a general functional relation between traditionally defined Q and FBW. The proof is based on an analytical evaluation of the functional relation for two distinct RLC circuits. It is given by contradiction, and it also covers some other commonly used prescriptions of Q. Second, it is pointed out that the so-called $\QZ$ factor defined in \cite{KajfezWheless_InvariantDefinitionsOfTheUnloadedQfactor} and further generalized in \cite{YaghjianBest_ImpedanceBandwidthAndQOfAntennas} is inversely proportional to FBW for first order systems, but ceases to have this behaviour for higher order systems.

%%%%%%%%%%%%%%%%%%%%%%%%%%%%%%%%%%%%%%%%%%%%%%%%%%%%%%%%%%%%%%%%%%%%%%%%%%%%%%%%%%%%%%%%%%%%%%%%%%%%%%%%%%%%%%%%%%%%%%%%%%%%%%%%%%%%%%%%%%%%%%%%%%%%%%%%%%%%%%%%%%%%%%%%%%%%%%%%
\section{Definition of the Q factor}
\label{theory}

This Section defines several widely used prescriptions of the Q factor that will be used later:
\begin{itemize}
\item classical $\Qc$, \cite{IEEEStd_antennas},
\item modified $\Qrev$, based on the concept of reversible energy, \cite{GrimesEtAl_TimeDomainMeasurementOfAntennaQ},
\item $\QX$, based on differentiation of the input reactance, \cite{Rhodes_ObservableStoredEnergiesOfElectromagneticSystems},
\item $\QZ$, based on differentiation of the input impedance, \cite{KajfezWheless_InvariantDefinitionsOfTheUnloadedQfactor, YaghjianBest_ImpedanceBandwidthAndQOfAntennas}.
\end{itemize}

%----------------------------------------------------------
\subsection{Classical definition of the Q factor}
\label{Q1}

The classical Q factor is conventionally defined as \cite{IEEEStd_antennas} 
\BE
\label{Qtopic_1A}
\Qc = \frac{\omega_0 \Wst}{\PL},
\EE
in which $\omega_0$ is the resonant frequency, $\Wst$ is the cycle mean stored energy, and $\PL$ is the cycle mean power loss. This prescription of the Q factor is traditionally encumbered with difficulties in identifying of the stored energy of a general electromagnetic radiator \cite{Vandenbosch_Reply2Comments}. This problem is however left aside in this note, as $\Wst$ is used only for non-radiating circuits for which the concept of stored energy is well established \cite{Hallen_ElectromagneticTheory}. Namely, the cycle mean stored energy of a non-radiating circuit can generally be written as
\BE
\label{QvsFBW_10}
\Wst = \frac{1}{4} \sum\limits_n \left(  L_n \left|I_{\mathrm{L}_n}\right|^2 + C_n \left|U_{\mathrm{C}_n}\right|^2 \right),
\EE
and the cycle mean lost power can be written as
\BE
\label{QvsFBW_11}
\PL = \frac{1}{2} \sum\limits_n R_n \left|I_{\mathrm{R}_n}\right|^2.
\EE
In (\ref{QvsFBW_10}) and (\ref{QvsFBW_11}), $L$, $C$, $R$ are the inductance, capacitance, and resistance of the circuit, and $I_\mathrm{L}$, $U_\mathrm{C}$, $I_\mathrm{R}$ are the corresponding currents and voltages. The convention \mbox{$\mathcal{F}\left(t\right) = \Re\left\{F\left(\omega\right) \mathrm{exp}(\jmath\omega t)\right\}$} for time-harmonic quantities has been utilized.
%Even thought the formula (\ref{Qtopic_1A}) is formally extremely simple to read, it is still not fully understood what is the true stored energy. There are some recent efforts how correctly define the stored energy in time domain \cite{CapekJelinekVandenboschHazdra_EMenergiesAndRadiationQfactor}, however evaluation of related expressions is extremely involved even for the basic radiators like dipole. Moreover, it will be proved in the following Section~\ref{propor} that (\ref{Qtopic_1A}) is not uniquely proportional to the fractional bandwidth (\ref{FBW_1A}), which violates the introductory premise (\ref{FBW_0A}).

%----------------------------------------------------------
\subsection{$\Qrev$ factor based on reversible energy}
\label{Q2}

The original definition of Q (\ref{Qtopic_1A}) can be slightly modified to
\BE
\label{Qtopic_2A}
\Qrev = \frac{\omega_0 \Wrev}{\PL},
\EE
in which $\Wrev$ (so-called reversible energy) is that part of the stored energy $\Wst$ which can be recovered back from the input port by a matched load. This reversible energy can in essence be evaluated by bringing the system into a time-harmonic steady state at frequency $\omega_0$ by a voltage source with matched internal impedance, afterwards switching off the source and capturing all the energy returned to the internal impedance \cite{GrimesEtAl_TimeDomainMeasurementOfAntennaQ}. A detailed description of this method for a general radiator can be found in \cite{CapekJelinekVandenboschHazdra_EMenergiesAndRadiationQfactor}.

%----------------------------------------------------------
\subsection{Reactance $\QX$ factor}
\label{Q3}

A different approach to defining Q is based on the assumption that Foster's reactance theorem \cite{Foster_AreactanceTheorem} also holds for lossy systems \cite{Rhodes_AReactanceTheorem, GeyiJarmuszewskiQi_TheFosterReactanceTheoremForAntenansAndRadiationQ}. In that case, Q can be defined by the frequency derivative of the input reactance as
\BE
\label{Qtopic_3A}
\QX = \frac{\omega_0}{2 \RE{\Zin}} \left|\frac{\partial \IM{\Zin}}{\partial\omega}\right|_{\omega = \omega_0},
\EE
where $\Zin$ is the input impedance of the circuits. This definition was proposed by Harrington \cite{HarringtonMautz_ControlOfRadarScatteringByReactiveLoading}, and was refined by Rhodes \cite{ Rhodes_ObservableStoredEnergiesOfElectromagneticSystems}, and it is commonly used even nowadays. 

% However, since it is widely accepted that the Foster theorem does not hold for real antennas \cite{Best_TheFosterReactanceTheoremAndQualityFactorForAntennas}, the definition (\ref{Qtopic_3A}) is sufficient only in vicinity of the resonances \cite{CapekJelinekHazdraEichler_MeasurableQ}.

%----------------------------------------------------------
\subsection{Impedance $\QZ$ factor}
\label{Q4}

A prescription which is widely used in antenna practice gives the Q factor in terms of the input impedance \cite{KajfezWheless_InvariantDefinitionsOfTheUnloadedQfactor, YaghjianBest_ImpedanceBandwidthAndQOfAntennas}. The relation reads
\BE
\label{Qtopic_4A}
\QZ = \frac{\omega_0}{2 \RE{\Zin}} \left|\frac{\partial \Zin}{\partial\omega}\right|_{\omega = \omega_0}
\EE
and it is known to correspond well to FBW \cite{YaghjianBest_ImpedanceBandwidthAndQOfAntennas}.
%The $\QZ$ is proportional to the VSWR bandwidth \cite{YaghjianBest_ImpedanceBandwidthAndQOfAntennas}, and is clearly based on measurable quantity (input impedance $\Zin$). In addition, it will be proved that (\ref{Qtopic_4A}) is exactly proportional to the FBW, and thus represents a consistent definition of Q.

%%%%%%%%%%%%%%%%%%%%%%%%%%%%%%%%%%%%%%%%%%%%%%%%%%%%%%%%%%%%%%%%%%%%%%%%%%%%%%%%%%%%%%%%%%%%%%%%%%%%%%%%%%%%%%%%%%%%%%%%%%%%%%%%%%%%%%%%%%%%%%%%%%%%%%%%%%%%%%%%%%%%%%%%%%%%%%%
\section{Functional Relation of Q and FBW}
\label{propor}

The major purpose of this note is to investigate the functional relation 
\BE
\label{FBW_0A}
\mathrm{FBW} = f \left( Q \right),
\EE
where $f$ is an as yet unknown function and
\BE
\label{FBW_1A}
\mathrm{FBW} = \frac{\omega_+ - \omega_-}{\omega_0},
\EE
in which $\omega_+$ and $\omega_-$ delimit the operational range of an antenna. 

%----------------------------------------------------------
\subsection{First-order systems}
\label{FirstOrder}

Instead of directly analysing relationship (\ref{FBW_0A}) for a complex system such as an antenna, we start with two single-resonant RLC circuits, depicted in Fig.~\ref{fig:RLC1}. 
\BF
\includegraphics[width=8.9cm]{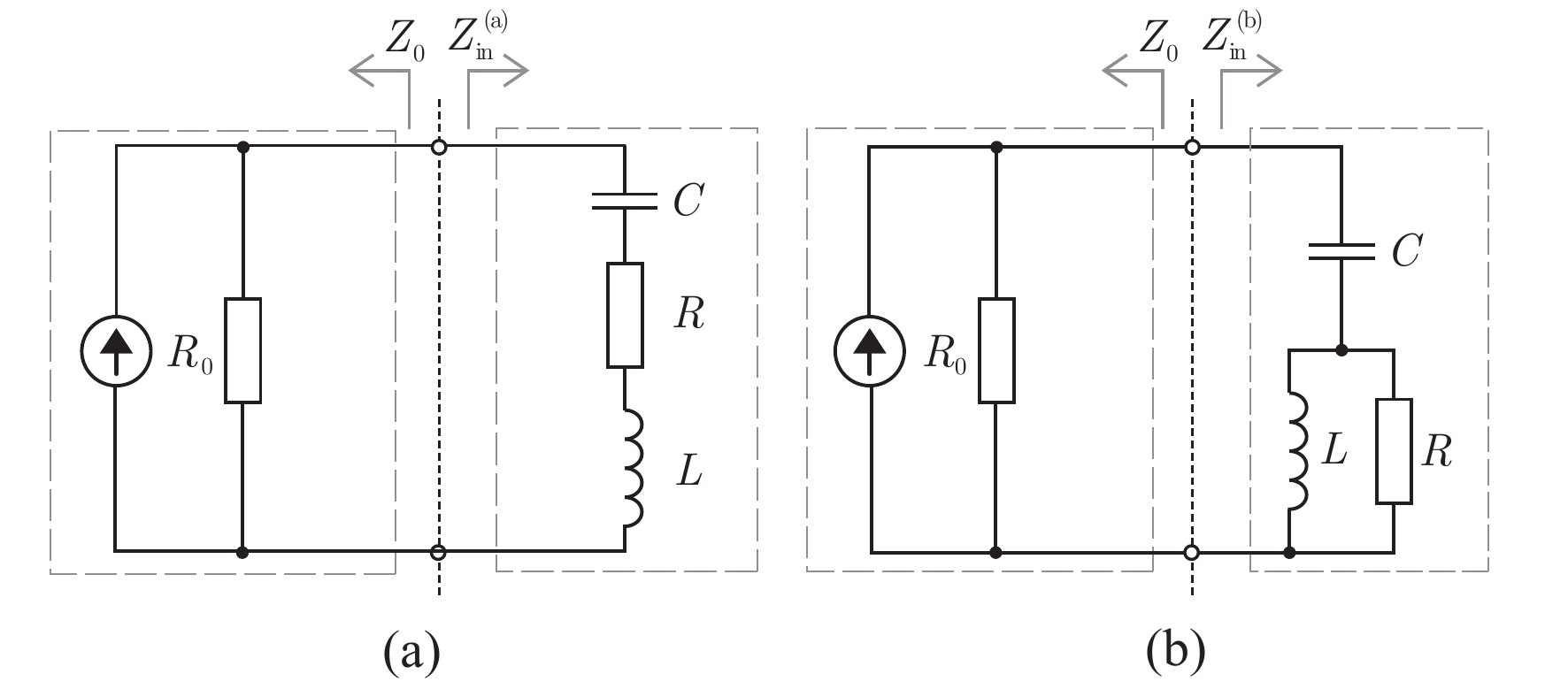}
\caption{The studied RLC circuits connected to a voltage source with internal resistance $R_0$: (a) $R$, $C$, and $L$ in series, (b) $C$ in series with parallel $L$ and $R$.}
\label{fig:RLC1}
\EF
If it is proved that a given definition of Q is not uniquely proportional to FBW for the simple circuits in Fig.~\ref{fig:RLC1}, it can be concluded that this Q is not proportional to FBW at all.
\BF
\includegraphics[width=7cm]{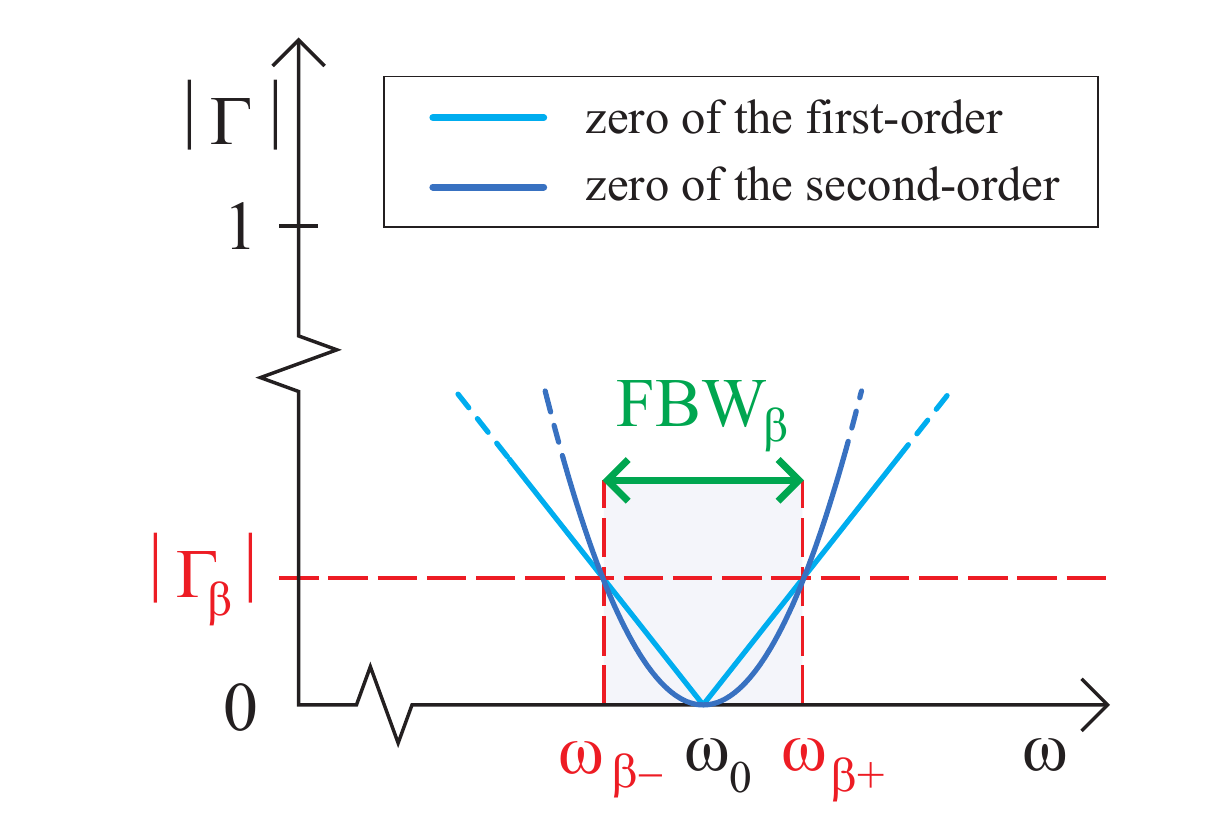}
\caption{The course of the reflectance in the vicinity of the circuit's resonance $\omega_0$. The circuit is assumed to be matched ($Z_\mathrm{in} \left( \omega_0 \right) = R_0$) at resonance frequency $\omega_0$, so that $\left| \Gamma \left(\omega_0\right)\right| = 0$.}
\label{fig:FBW0}
\EF

We utilize a simple consideration depicted in Fig.~\ref{fig:FBW0}, which assumes that the reflectance
\BE
\label{FBW_6A}
\left|\Gamma\right| = \left| \frac{Z_\mathrm{in} - Z_0}{Z_\mathrm{in} + Z_0}\right|
\EE
can be expanded to its Taylor series and only the first non-zero term can be kept near the circuit's resonance. Without loss of generality, we also assume that the circuit is matched to the input resistance (\mbox{$\Zin\left(\omega_0\right)=R_0$}). Under such conditions the reflectance can be written as
\BE
\label{QvsFBW_0}
\begin{split}
\left|\Gamma\right| &= \left| \omega - \omega_0 \right| \left| \frac{\partial\Gamma}{\partial\omega}\right|_{\omega = \omega_0} + \mathcal{O}\left(\omega^2\right) \\
&\approx \frac{\left|\omega - \omega_0\right|}{\omega_0} \frac{\omega_0}{2 \Re\{\Zin\}} \left|\frac{\partial \Zin}{\partial\omega}\right|_{\omega = \omega_0}= \frac{\mathrm{FBW}}{2} \QZ,
\end{split}
\EE
which for $\left|\Gamma_\beta\right| \rightarrow 0$ gives the required functional relation (\ref{FBW_0A})
\BE
\label{QvsFBW_1}
\mathrm{FBW}_\beta = 2 \frac{\left| \Gamma_\beta \right|}{\QZ}.
\EE
This means that the $\QZ$ factor (\ref{Qtopic_4A}) is uniquely proportional to the FBW (at least in this differential sense). Furthermore, if the other Q factors are to follow relation (\ref{FBW_0A}) they must necessarily be functionally dependent on $\QZ$. This property is investigated in the following.

The $\QZ$ factors of the two circuits under consideration can easily be calculated from the input impedances, which yields
\begin{subequations}
\begin{alignat}{2}
\label{QvsFBW_1A}
\QZ^{(\mathrm{a})} &= \frac{\omega_0^{\mathrm{(a)}} L}{R}, && \omega_0^\mathrm{(a)} = \frac{1}{\displaystyle\sqrt{L C}},\\ 
\label{QvsFBW_1B}
\QZ^{(\mathrm{b})} &= \frac{R}{\omega_0^{\mathrm{(b)}} L} \frac{1}{\omega_0 \sqrt{L C}}, \quad && \omega_0^\mathrm{(b)} = \frac{R}{L \sqrt{\displaystyle\frac{C R^2}{L} - 1}},
\end{alignat}
\end{subequations}
where superscripts (a) and (b) refer to the two circuits in Fig.~\ref{fig:RLC1}. The other Q factors can be evaluated in a straightforward manner as
\begin{subequations}
\begin{align}
\label{QvsFBW_2A}
\Qc^{(\mathrm{a})} &= \QZ^{(\mathrm{a})},\\ 
\label{QvsFBW_2B}
\QX^{(\mathrm{a})} &= \QZ^{(\mathrm{a})},\\ 
\label{QvsFBW_2C}
\Qrev^{(\mathrm{a})} &= \frac{\QZ^{(\mathrm{a})}}{2},
\end{align}
\end{subequations}
and
\begin{subequations}
\begin{align}
\label{QvsFBW_3A}
\Qc^{(\mathrm{b})} &= \xi_\mathrm{cl} \, \QZ^{(\mathrm{b})},\\ 
\label{QvsFBW_3B}
\QX^{(\mathrm{b})} &= \xi_X \, \QZ^{(\mathrm{b})},\\
\label{QvsFBW_3C}
\Qrev^{(\mathrm{b})} &= \xi_\mathrm{rev} \, \frac{\QZ^{(\mathrm{b})}}{2},
\end{align}
\end{subequations}
where
\begin{subequations}
\begin{align}
\label{QvsFBW_4A}
\xi_\mathrm{cl} &= \sqrt{\chi},\\ 
\label{QvsFBW_4B}
\xi_X &= \sqrt{\chi}\,\frac{\chi}{\chi + \displaystyle \left(\QZ^{(\mathrm{b})} \right)^{-2}},\\ 
\label{QvsFBW_4C}
\xi_\mathrm{rev} &= \sqrt{\chi} \, \frac{\chi + \displaystyle \left(\QZ^{(\mathrm{b})} \right)^{-2}}{\chi + \displaystyle 2 \left(\QZ^{(\mathrm{b})} \right)^{-2}},\\
\label{QvsFBW_4D}
\chi &= \frac{1 + \sqrt{1 + \displaystyle 4 \left(\QZ^{(\mathrm{b})} \right)^{-2}}}{2}. 
\end{align}
\end{subequations}

The above results offer a simple interpretation. Since the $\QZ$ factor has been shown to have a unique functional relation to FBW, see (\ref{QvsFBW_1}), the other quality factors could have such a unique functional relation only if the functional relations corresponding to circuit (a) and circuit (b), see (13) and (14), are the same, i.e. if the corresponding $\xi$ coefficients in (\ref{QvsFBW_4A}), (\ref{QvsFBW_4B}), (\ref{QvsFBW_4C}) are equal to unity. That this is not the case is clear from their analytical prescription and also from their graphical representation in Fig.~\ref{fig:RLC2}. 
\BF
\includegraphics[width=8.9cm]{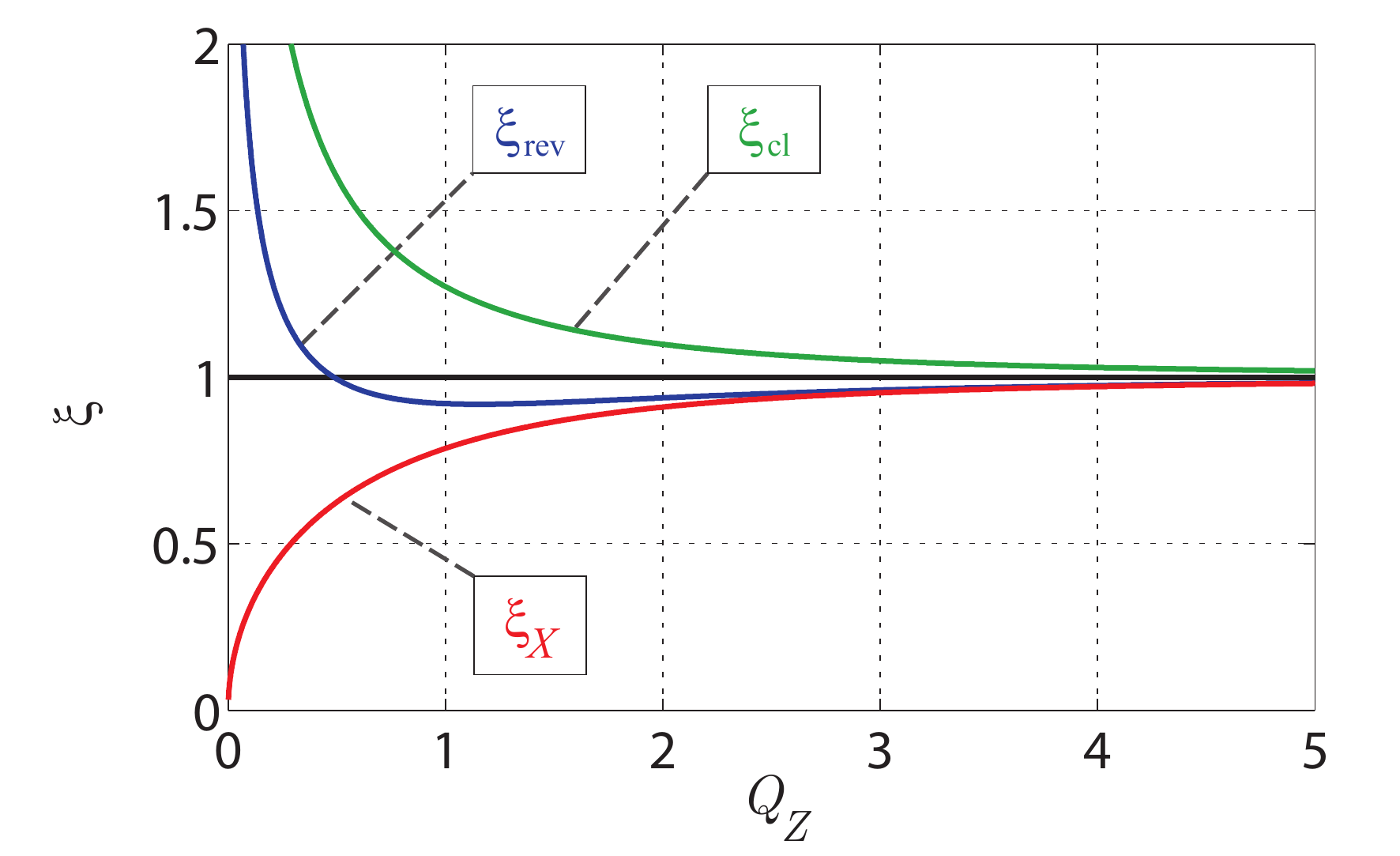}
\caption{The $\xi$ factors of (\ref{QvsFBW_4A})--(\ref{QvsFBW_4C}) as a function of $\QZ$.}
\label{fig:RLC2}
\EF
By means of contradiction, it must then be stated that there is no general functional relation between FBW and $\Qc,\QX,\Qrev$, the only exception being $Q \rightarrow \infty$.

%----------------------------------------------------------
\subsection{Higher-order systems}
\label{SecondOrder}

Section~\ref{FirstOrder} has shown that, in the case of first-order systems, only $Q_Z$ is a potential candidate for having a general functional relation to FBW. The purpose of this subsection is to test this property on higher-order systems.

Higher-order systems offer more degrees of freedom. This in general makes approximation (\ref{QvsFBW_0}) invalid. In fact, it can be shown \cite{Collin_FoundationsForMicrowaveEngineering} that a circuit of order $n$ can always be tuned so that first $n-1$ terms of the Taylor expansion vanish (binomial transformer).

An example of such a second order system \cite{Gustaffson_StoredElectromagneticEnergy_arXiv, GustafssonNordebo_BandwidthQFactorAndResonanceModelsOfAntennas} is depicted in Fig.~\ref{fig:circuit2}, which for $K_\mathrm{P}=K_\mathrm{S}$ results in $Q_Z=0$. 
\BF
\includegraphics[width=8.9cm]{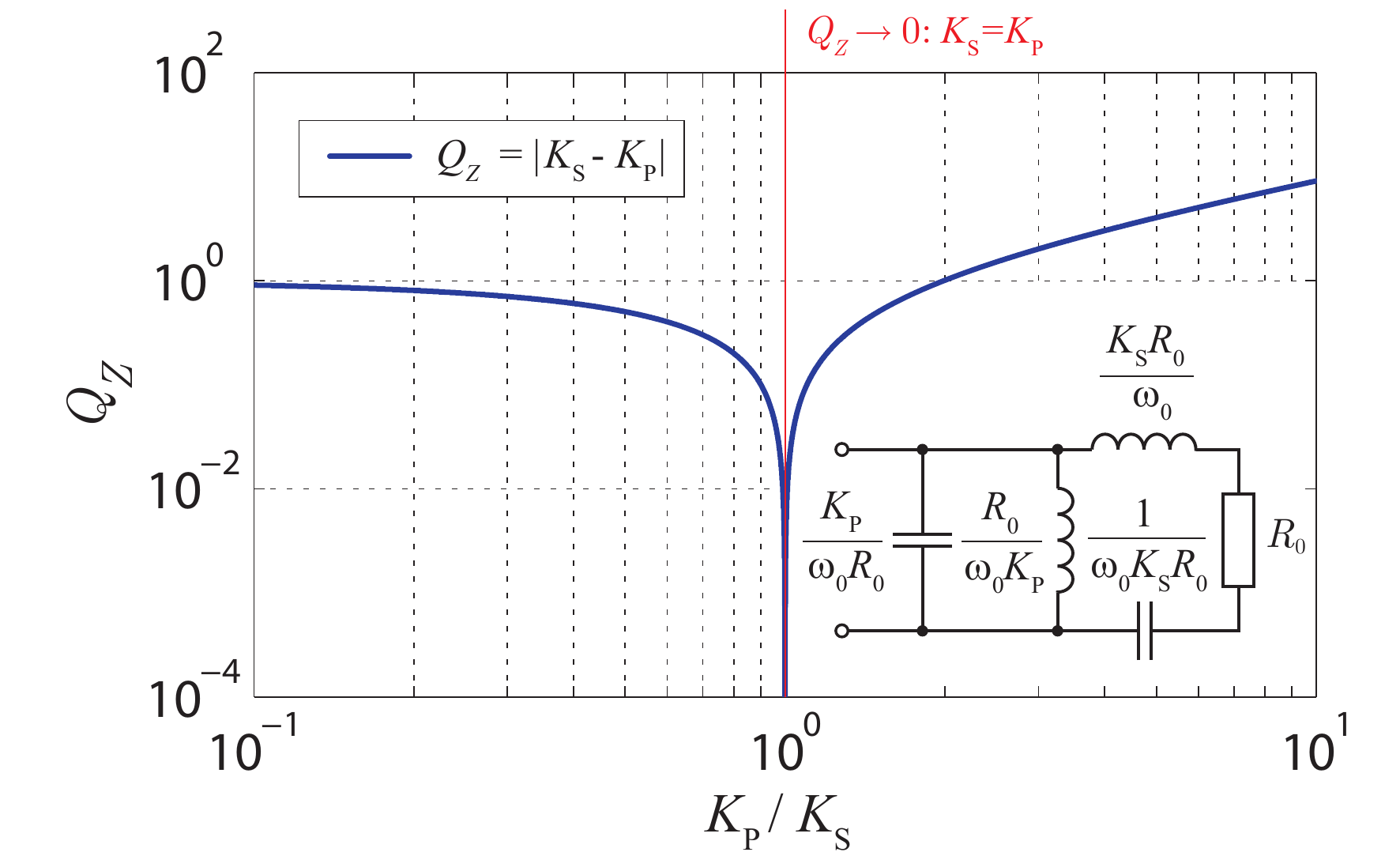}
\caption{The quality factor $Q_Z$ of a selected second-order RLC circuit.}
\label{fig:circuit2}
\EF
Another example \cite{Gustaffson_QdisperssiveMedia_arXiv} is a thin-strip dipole of length $L$ and width \mbox{$w = L / 100$}, see Fig.~\ref{fig:dipole1}. If the dipole is fed by a voltage gap placed at $h \approx 0.228 L$, we realize that  $Q_Z = 0$ at $ka \approx 6.171$, in which $k = \omega / c_0$ is the wavenumber, $c_0$ is the speed of light, and $a$ is the radius of the smallest circumscribing sphere.
\BF
\includegraphics[width=8.9cm]{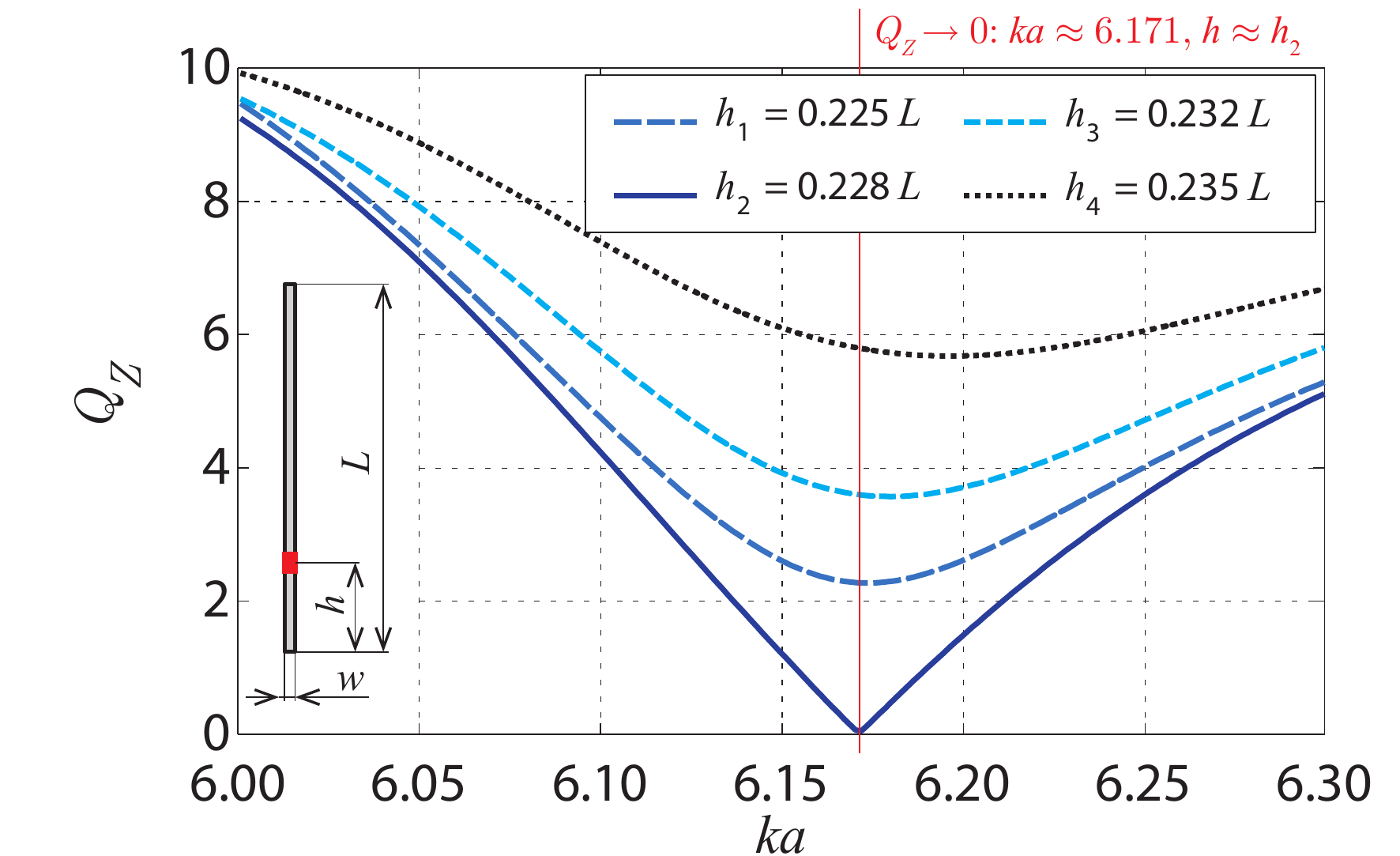}
\caption{The quality factor $Q_Z$ of a thin-strip dipole as a function of electrical size and as a function of feeding position.}
\label{fig:dipole1}
\EF

This awkward property of a possibly zero value of $Q_Z$ in the case of circuits with clearly finite FBW unfortunately exclude $Q_Z$ from prescriptions with a possibly unique relation to FBW.

\section{Conclusion}
\label{concl}

It has been shown that, contrary to common belief, the classical quality factor defined by the stored and lost energy is not related to the fractional bandwidth by a general and unambiguous functional relation. This is also true for Q factors resulting from reversible energy and input reactance. Considering the first-order system, only the Q factor based on differentiation of the input impedance has been shown to be a possible candidate for such a general functional relation. It has however been demonstrated that for higher-order systems, including elementary radiators like dipoles, no quality factor has in general exact proportionality to the fractional bandwidth.

%Under these circumstances, the presence of classical Q definition in IEEE Standard is questionable and leads to wrong expectations since this Q is in fact not a legitimate definition.

\section*{Acknowledgement}
The authors would like to thank to Mats Gustafsson from Lund University, Sweden for a fruitful discussion about the topic, and for pointing out the possibility of minimizing $Q_Z$ by off-center feeding of a dipole.

\ifCLASSOPTIONcaptionsoff
  \newpage
\fi

\bibliographystyle{IEEEtran}
\bibliography{references_LIST}

% =========================================================================================================
\end{document}